\documentclass[conference]{IEEEtran}

\usepackage{amsmath}
\usepackage{amsfonts}
\usepackage{graphicx}
\usepackage{color}
\usepackage[short]{optidef}
\usepackage{newtxtext,newtxmath} 
\usepackage{mleftright} 
\mleftright 
\usepackage[margin=0.70in]{geometry}
\usepackage{cite}
\newcommand{\Fourier}{\mbox{\boldmath$\cal F$}}
\newcommand {\blambda} {\mbox{\boldmath{$\lambda$}}}

\newcommand {\bpsi} {\mbox{\boldmath{$\psi$}}}

\newcounter{temp_count} 

\IEEEoverridecommandlockouts
\begin{document}

\title{Resource Allocation for Single Carrier Massive MIMO Systems}
\date{2 February 2022}
 \author{%
    \IEEEauthorblockN{Brent A. Kenney\IEEEauthorrefmark{1}, Arslan J. Majid\IEEEauthorrefmark{2}, Hussein Moradi\IEEEauthorrefmark{2}, and Behrouz Farhang-Boroujeny\IEEEauthorrefmark{1}}
    \IEEEauthorblockA{\IEEEauthorrefmark{1}Electrical and Computer Engineering Department, University of Utah, Salt Lake City, Utah, USA}
    \IEEEauthorblockA{\IEEEauthorrefmark{2}Idaho National Laboratory, Salt Lake City, Utah, USA\\}
\thanks{This manuscript has in part been authored by Battelle Energy Alliance, LLC under Contract No. DE-AC07-05ID14517 with the U.S. Department of Energy. The United States Government retains and the publisher, by accepting the paper for publication, acknowledges that the United States Government retains a nonexclusive, paid-up, irrevocable, world-wide license to publish or reproduce the published form of this manuscript, or allow others to do so, for United States Government purposes. \bf{STI Number: INL/CON-21-64819}.}
}

\maketitle
\begin{abstract}
Resource allocation in orthogonal frequency division multiplexing (OFDM) systems is performed through allocating blocks of subcarriers to each user. Even though OFDM is the primary waveform for 5G NR systems, research reports have noted that single carrier modulation (SCM) offers several advantages over OFDM in massive multiple input multiple output (MIMO) systems, making it a preferred candidate for some future applications such as massive machine type communications (mMTC).  This paper presents a method for SCM resource allocation and the relevant information recovery algorithms at the receiver.  Our emphasis is on cyclic prefixed SCM, where highly flexible and efficient frequency domain detection algorithms enable the operation of many simultaneous users in a massive MIMO uplink scenario. The proposed resource allocation method allows the number of users to exceed the number of antennas at the base station (BS).  Each single carrier transmission is partitioned into $L$ interleaved streams, and each user is allocated a number of such streams.  One major benefit of SCM is that each data symbol is spread over the entire bandwidth. As such, the receiver performance is dictated by the average channel gain across the transmission band rather than the channel gain at a given frequency bin or a small group of frequencies.  In the proposed setup, each stream may be thought of as a resource block in SCM, analogous to resource blocks in OFDM. Hence, in the context of this paper, the terms resource blocks and streams may be used interchangeably.
\end{abstract}

\section{Introduction}
Orthogonal frequency division multiplexing (OFDM) is the dominant waveform for today's massive multiple input multiple output (MIMO) systems.  One reason for OFDM's widespread usage is its flexible resource allocation.  Resource blocks are assigned as groups of subcarriers, thereby dividing the available spectrum between several users.  While OFDM enjoys these advantages, it also has its drawbacks, which makes single carrier modulation (SCM) a better choice in some instances.  For example, SCM is more tolerant to carrier frequency offset, Doppler shifts, and channel aging \cite{Wang:2004}, \cite{Wu:2016}, \cite{Mokhtari:2020}.  

SCM is a promising alternative solution for massive machine type communication (mMTC), which is characterized by low data rates and densely spaced terminals.  Hence, it is critical for the waveform to allocate resources to a large number of simultaneous users properly.  In this paper, we introduce a processing method to provide flexible resource allocation using SCM.  The proposed method is capable of simultaneously servicing more user equipments (UEs) than there are base station (BS) antennas, similar to OFDM.  In addition, the massive MIMO processing gain is maintained at a high level that is commensurate with the total symbol rate from all UEs rather than being based on the number of UEs.  Because of the high processing gain and the ability to service several simultaneous users in a resource-agile manner, this method is the full-bandwidth SCM analogue to resource blocks in traditional OFDM-based systems.  This is in contrast to a partial-band SCM method such as Discrete Fourier Transform (DFT) precoded OFDM.  To the best of our knowledge, there is no prior work in the literature that discusses resource allocation for full-bandwidth SCM.

An efficient method of detection for uplink (UL) operation with cyclic prefixed SCM (CP-SCM) was reported in \cite{Kenney:TWC:2021} for the case where the number of BS antennas was greater than the number of single-antenna users (i.e., $M>K$).  The solution in \cite{Kenney:TWC:2021} assumed that each UE was transmitting $N$ symbols per frame, where $N$ is the number of samples per signal frame, excluding the cyclic prefix (CP).  $N$ is also the equivalent number of frequency bins in the frequency domain (FD).  In this paper we show that in the event that UEs have less than $N$ symbols to transmit per frame, the available time-frequency resources can be allocated differently in order to service a greater number of simultaneous UEs (e.g., $K>M$).  In order to support a variable degree of service per UE and increase the number of potential UEs, the data symbols from each UE are rearranged to form between $1$ and $L$ interleaved data streams.  

The Cyclic Prefix Direct Sequence Spread Spectrum (CP-DSSS) waveform analyzed in \cite{Kenney:CPDSSS:2020} is an example of a CP-SCM waveform that sends a single data stream and uses Zadoff-Chu (ZC) sequences for spreading.  Even though the unused data streams result in gaps between symbols, the spreading sequence redistributes the energy of each symbol, resulting in a more power-efficient waveform.  Here, we expand the CP-DSSS definition to include multiple data streams.  This work also applies to other CP-SCM waveforms that arrange data symbols in the same manner.

The approach outlined in this paper is based on minimum mean-squared error (MMSE) detection and will be referred to as multi-stream processing (MSP).  Based on the resources allocated to the UE by the BS, an active UE may populate one or more of these streams with data in a given UL interval, which is equivalent to one OFDM symbol interval.  
One of the major advantages of MSP is the superior massive MIMO processing gain.  For cases where the number of active streams is less than the maximum number of possible data streams, $L$, MSP produces a higher processing gain than the UL detection algorithm in \cite{Kenney:TWC:2021}.  The increased processing gain is due to the creation of virtual antennas in the MSP algorithm.  Virtual antennas are created when a single data stream is upsampled by $L$, which creates $L$ copies of the signal spectrum across the band.  Each of these signal copies can be used for processing, which results in $L$ virtual antennas for each physical antenna at the BS.  In the same way that additional antennas improve the processing gain of massive MIMO systems \cite{Rusek:2013}, virtual antennas also increase the processing gain.  The virtual antenna concept is more complicated when applied to multiple streams per UE, but it allows flexibility in how different amounts of resources are assigned to different users.  

The virtual antenna term has been used in the following publications \cite{Chen:2016}, \cite{Han:2016}, \cite{Xu:2018} in the context of time reversal (TR) precoding and scatterers in the environment, while transmitting from a single antenna.  Through TR precoding, the number of virtual antennas was defined by the number of taps in the channel impulse response.  Interestingly, the TR implementation of virtual antennas also required that the data be upsampled, but that upsampling was used to limit inter-symbol interference instead of creating spectral copies.  

This paper makes the following contributions to the area of CP-SCM processing for massive MIMO systems:
\begin{itemize}
\item Presents a novel solution for multi-user resource allocation using full-bandwidth CP-SCM;
\item Quantizes resources in terms of data streams that are a system-defined fraction of the available symbol rate (i.e., spectral resources);
\item Defines a means of detecting more single-stream users than there are physical BS antennas;
\item Expands the processing to handle multiple streams per user in a flexible manner to enable customized resource allocation;
\item Shows that the resulting processing gain is based on the number of BS antennas, $M$, and the aggregate user symbol rate rather than the number of users.
\end{itemize}

This paper is organized into the following sections: Section \ref{System_Model} introduces the framework in which the single-stream SCM transmission and detection is defined; Section \ref{Single_Stream_Detection} presents the approach of single-stream MSP communication; Section \ref{Multi_Stream_Detection} extends the approach to allow users to occupy multiple streams; and Section \ref{Conclusion} provides concluding remarks.

\textit{Notation:} Italic letters represent scalars.  Bold lowercase letters represent column vectors.  FD vectors are capped with a tilde.  Bold uppercase letters represent matrices.  $\mathbf{I}_K$ is the $K \times K$ identity matrix.  $(\cdot )^*$, $(\cdot )^{\textrm{T}}$, and $(\cdot )^{\textrm{H}}$ represent the complex conjugate, transpose, and Hermitian operators, respectively.  $\mathbb{E}[\cdot ]$ is the expected value taken over all channel and noise realizations.  Finally, $\textrm{tr}\{\cdot \}$ is the trace operator.

\section{System Model for Single-stream SCM} \label{System_Model}
The scenarios modeled in this paper assume that the UE does not have any channel state information (CSI).  Each UE is assumed to have knowledge of the average channel power so that it can maintain a power target at the BS, averaged across all BS antennas.  The BS has CSI between each UE and each of the $M$ BS antennas, which can be obtained by transmitting pilot signals from the UEs.  This paper assumes perfect CSI at the BS.

\begin{figure}[!t]
    \centering
    \includegraphics[width=3.4in, clip=true, trim=2.0cm 7.25cm 1.5cm 10.5cm ]{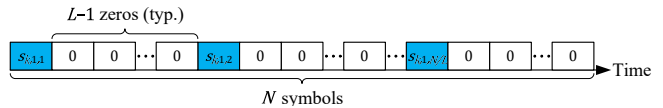}
    \caption{Single-stream upsampled symbols.  $N/L$ symbols are transmitted with $L-1$ zeros inserted between each symbol.}
    \label{single_stream_upsampling}
\end{figure}

The UL transmission is divided into frames of length $N$ in which each user transmits $N/L$ unit variance symbols, represented as $\mathbf{s}_{k,1}^{\textrm{UL}}$ for the $k^{\textrm{th}}$ UE, where the second index with a value of $1$ indicates that it is the first stream.  The $N/L$ symbols are equally spaced with $L-1$ zeros between symbols as shown in Fig. \ref{single_stream_upsampling}.  A spreading sequence is applied to fill in the gaps between symbols.  A CP is added to the front of the frame prior to transmission to preserve circular convolution.  The transmitted sequence from the $k^{\textrm{th}}$ UE before addition of the CP is given as
\begin{equation}
    \mathbf{x}_k^{\textrm{UL}} =  \mathbf{Z} \mathbf{E}_L \mathbf{s}_{k,1}^{\textrm{UL}},
    \label{eq:01}
\end{equation}
where $\mathbf{x}_k^{\textrm{UL}}$ is a length-$N$ vector, $\mathbf{Z}$ is the $N \times N$ spreading matrix, and $\mathbf{E}_L$ is the expander matrix that converts the $N/L$ unit-variance symbols in $\mathbf{s}_{k,1}$ to a length-$N$ vector.  Here, we assume that $\mathbf{Z}$ is a unitary matrix (i.e., $\mathbf{Z}^{\textrm{H}} \mathbf{Z} = \mathbf{I}_N$).  Consequently, $\mathbf{Z}$ does not alter the transmit power.  In addition, applying $\mathbf{Z}^{\textrm{H}}$ at the receiver does not alter the noise statistics.  As in the case of CP-DSSS, we also assume that $\mathbf{Z}$ is circulant.  This property will prove useful for efficient despreading in the FD.  The expander matrix, $\mathbf{E}_L$, performs the upsample function and is formed by taking $\mathbf{I}_{N/L}$ and inserting $L-1$ rows of zeros after each row, resulting in a matrix with dimensions $N \times (N/L)$.

The received signal from the $K$ UEs at the $m^{\textrm{th}}$ antenna after CP removal is expressed  as
\begin{equation}
    \mathbf{y}_m^{\textrm{UL}} = \sum_{ k=1}^K \mathbf{H}_{m,k} \mathbf{x}_k^{\textrm{UL}}  + \mathbf{w}_m, 
    \label{eq:02}
\end{equation}
where $\mathbf{H}_{m,k}$ is the $N \times N$ circulant convolutional channel matrix between antenna $m$ and user $k$ and $\mathbf{w}_m$ is the receiver noise ($\mathbf{w}_m \sim \mathcal{CN}( \mathbf{0}, \sigma^2_w \mathbf{I}_N )$) at the $m^{\textrm{th}}$ antenna.  

Let $\mathbf{h}_{m,k}$ represent the channel impulse response vector between antenna $m$ and user $k$, which is of length $L_h$.  $\mathbf{H}_{m,k}$ is formed by first taking $\mathbf{h}_{m,k}$ and appending $N-L_h$ zeros to form the base vector ${\mathbf{h}_{m,k} }_{(0)}$.  We then take downward cyclic shifts of ${ \mathbf{h}_{m,k} }_{(0)}$ to create
\begin{equation}
  \mathbf{H}_{m,k}=
  \begingroup 
    \setlength\arraycolsep{4pt}
    \begin{bmatrix}
       { \mathbf{h}_{m,k} }_{(0)} & { \mathbf{h}_{m,k} }_{(1)} \dots & { \mathbf{h}_{m,k} }_{(N-2)} & { \mathbf{h}_{m,k} }_{(N-1)}
    \end{bmatrix}
  \endgroup, 
    \label{eq:03}
\end{equation}
where the parenthetical subscript represents the number of cyclical shifts applied to the base vector.  We will also use the parenthetical subscript later in the paper to cyclically shift the expander matrix, $\mathbf{E}_L$.  

In order to process the received signal in the FD, we take the $N$-point DFT of the received signal vector in \eqref{eq:02} after substituting \eqref{eq:01}.  This results in the expression
\begin{equation}
    \tilde{ \mathbf{y} }_m^{\textrm{UL}} = \sum_{k=1}^K \mathbf{\Lambda}_{m,k} \mathbf{\Omega} \frac{ 1 }{ \sqrt{L} } \mathbf{T}_{N,L} \tilde{ \mathbf{s} }_{k,1}^{\textrm{UL}} + \tilde{ \mathbf{w} }_m, 
    \label{eq:04}
\end{equation}
where the tilde represents the FD representation of the vectors and $\mathbf{\Lambda}_{m,k}$, $\mathbf{\Omega}$, and $ \mathbf{T}_{N,L} / \sqrt{L}$ are the FD representations of $\mathbf{H}_{m,k}$, $\mathbf{Z}$, and $\mathbf{E}_L$, respectively.  The FD representation of a matrix may not be a common term, but it is applicable in cases where a matrix can be decomposed using a DFT matrix and an inverse DFT (IDFT) matrix.  It is helpful to realize that the DFT decomposition is similar to a singular value decomposition (SVD).  In fact, the two are related for circulant matrices since the absolute values of the FD representation elements are equal to the singular values of the SVD \cite{Gray:2006}.

Any circulant $N \times N$ matrix (e.g., $\mathbf{H}_{m,k}$ and $\mathbf{Z}$) is diagonalized by the $N$-point DFT matrix,  $\Fourier_N$.  Consequently, the $\mathbf{\Lambda}_{m,k}$ and $\mathbf{\Omega}$ matricies are diagonal.  We note that $\Fourier_N$ is scaled such that $\Fourier_N^{-1} = \Fourier_N^{\textrm{H}}$.  Hence, $\mathbf{H}_{m,k} = \Fourier_N^{-1} \mathbf{\Lambda}_{m,k} \Fourier_N$ and $\mathbf{Z} = \Fourier_N^{-1} \mathbf{\Omega} \Fourier_N$ \cite{Gray:2006}.  It follows that taking the $N$-point DFT of the columns of $\mathbf{H}_{m,k}$ results in $\Fourier_N \mathbf{H}_{m,k} = \mathbf{\Lambda}_{m,k} \Fourier_N$, where $\mathbf{\Lambda}_{m,k}$ is a diagonal matrix containing the eigenvalues of $\mathbf{H}_{m,k}$.  Let $\lambda_{m,k,i}$ represent the $i^{\textrm{th}}$ value along the diagonal of $\mathbf{\Lambda}_{m,k}$.  The eigenvalues can also be obtained by taking the N-point DFT of the channel impulse response $\mathbf{h}_{m,k}$, which is used to form $\mathbf{H}_{m,k}$.  For more efficient computation, it is noted that all of the FD conversions can be performed with the Fast Fourier Transform (FFT) instead of the DFT.  We use the fact that $\mathbf{\Omega}$ is diagonal to show that $\mathbf{\Lambda}_{m,k}$ and $\mathbf{\Omega}$ are commutable.  As a result, despreading the received signal is performed in the FD by multiplying the received signal by $\mathbf{\Omega}^*$ since $\mathbf{\Omega}^* \mathbf{\Omega} =  \mathbf{I}_N$ because $\mathbf{Z}$ is unitary.  Based on this result, future uses of \eqref{eq:04} will drop the $\mathbf{\Omega}$ spreading term in order to focus on the processing after despreading.

The expander matrix, $\mathbf{E}_L$, is not circulant, but due to its structure, it can still be factored with the DFT and IDFT matrices.  Because $\mathbf{E}_L$ is not a square matrix, the size of the DFT and IDFT matrices differ, i.e., $\mathbf{E}_L = \Fourier_N^{-1} ( \mathbf{T}_{N,L} / \sqrt{L} ) \Fourier_{N/L}$, where $\Fourier_{N/L}$ is the $N/L$-point DFT matrix.  The matrix $\mathbf{T}_{N,L}$ is the $N \times (N/L)$ vertical tiling matrix, where $\mathbf{T}_{N,L} = [ \mathbf{I}_{N/L} \ \mathbf{I}_{N/L} \ \dots \ \mathbf{I}_{N/L} ]^{\textrm{T}}$.  We note here that $\tilde{\mathbf{s}}^{\textrm{UL}}_{k,1} = \Fourier_{N/L} \mathbf{s}^{\textrm{UL}}_{k,1}$.  Due to the tiling matrix structure, the FD representation of the $N/L$ symbols will be replicated $L$ times in the available spectrum with an amplitude scaling of $1/\sqrt{L}$.

\section{Single-stream Detection} \label{Single_Stream_Detection}
Single-stream detection in the FD follows a similar approach to the MRC-MMSE detector presented in \cite{Kenney:TWC:2021}.  As such, we define the vector of received signals for the $n^{\textrm{th}}$ bin as $\tilde{\mathbf{y}}_{:,n}^{\textrm{UL}} = [ \tilde{y}_{1,n}^{\textrm{UL}} \ \tilde{y}_{2,n}^{\textrm{UL}} \ \dots \ \tilde{y}_{M,n}^{\textrm{UL}} ]^{\textrm{T}}$.  One key observation for single-stream operation is that there are $L$ frequency bins of $\tilde{ \mathbf{y} }_m^{\textrm{UL}}$ that have components of $\tilde{s}_{k,1,n}^{\textrm{UL}}$, where the third index specifies the frequency bin.  Each of those bins are spaced by $N/L$ bins due to the structure of the tiling matrix, $\mathbf{T}_{N,L}$.  Hence, we define the composite vector of received signals as $\overline{\mathbf{y}}_{:,n}^{\textrm{UL}} = [ ( \tilde{\mathbf{y}}^{\textrm{UL}}_{:,n} )^{\textrm{T}} \ ( \tilde{\mathbf{y}}^{\textrm{UL}}_{:,n+N/L} )^{\textrm{T}} \ \dots \  ( \tilde{\mathbf{y}}^{\textrm{UL}}_{:,n+(L-1)N/L} )^{\textrm{T}} ]^{\textrm{T}}$, for $n=0, 1, ..., N/L-1$.

In order to express the equation for the composite received vector, $\overline{\mathbf{y}}_{:,n}^{\textrm{UL}}$, we now construct the  FD composite channel matrix.  The FD channel matrix for bin $n$ is based on the $n^{\textrm{th}}$ diagonal element of each $\mathbf{\Lambda}_{m,k}$ matrix and is defined as
\begin{equation}
	\mathbf{A}_n =
	\begin{bmatrix}
    	  \lambda_{1,1,n} & \lambda_{1,2,n} & \dots & \lambda_{1,K,n} \  \\
    	  \lambda_{2,1,n} & \lambda_{2,2,n} & \dots & \lambda_{2,K,n} \  \\
	  \vdots & \vdots & \ddots & \vdots \\
    	  \lambda_{M,1,n} & \lambda_{M,2,n} & \dots & \lambda_{M,K,n} \  \\
	\end{bmatrix}. 
	\label{eq:05}
\end{equation}

We note that the eigenvalues that form $\mathbf{A}_n$ are each a sum of the $L_h$ elements of the channel impulse response, $\mathbf{h}_{m,k}$, after being rotated according to the DFT coefficients.  Here, we assume that $\mathbf{h}_{m,k}$ is composed of zero-mean complex Gaussian random variables.  Since the channel gain is normalized to unity, the elements of $\mathbf{A}_n$ are also zero-mean complex Gaussian random variables with unit variance.  Given that the rows of $\mathbf{A}_n$ span all of the users and each row corresponds to a specific antenna, the rows of $\mathbf{A}_n$ are zero-mean, complex Gaussian vectors with covariance $\mathbf{I}_K$.  Consequently, the matrix $\mathbf{A}_n^{\textrm{H}} \mathbf{A}_n$ is a central complex Wishart matrix with $M$ degrees of freedom following Definition 2.3 of \cite{Verdu:Random_Matrix_Theory}.

The composite channel matrix for single-stream processing in the FD is defined as $\overline{\mathbf{A}}_n =[ \mathbf{A}_n^{\textrm{T}} \ \mathbf{A}_{n+N/L}^{\textrm{T}} \ \dots \ \mathbf{A}_{n+(L-1)N/L}^{\textrm{T}} ]^{\textrm{T}}$, for $n=0, 1, ..., N/L-1$.  Note that $\overline{\mathbf{A}}_n$ is an $ML \times K$ matrix.  Since each of the constituent channel matrices is taken from a different part of the spectrum, they appear as though they are from unique antennas.  By extension, the matrix $\overline{\mathbf{A}}_n^{\textrm{H}} \overline{\mathbf{A}}_n$ is a central complex Wishart matrix with $ML$ degrees of freedom.  In effect, we have created $ML$ virtual antennas from the $M$ physical antennas.  By increasing the number of virtual antennas by a factor of $L$, the single-stream processing has the ability to process a large number of UEs simultaneously.  In fact, the number of UEs can exceed the number of physical antennas as long as the following condition is met: $K < ML$.  The FD composite noise vector is defined as $\overline{\mathbf{w}}_{:,n} = [ ( \tilde{\mathbf{w}}_{:,n} )^{\textrm{T}} \ (\tilde{\mathbf{w}}_{:,n+N/L} )^{\textrm{T}} \ \dots \  ( \tilde{\mathbf{w}}_{:,n+(L-1)N/L} )^{\textrm{T}} ]^{\textrm{T}}$.  We can now express the FD composite received vectors for $n=0, 1, ..., N/L-1$ as
\begin{equation}
	\overline{\mathbf{y}}_{:,n}^{\textrm{UL}} = \frac{1}{\sqrt{L}} \overline{\mathbf{A}}_n \tilde{\mathbf{s}}_{:,1,n}^{\textrm{UL}} + \overline{\mathbf{w}}_{:,n}, 
	\label{eq:06}
\end{equation}
where $\tilde{\mathbf{s}}_{:,1,n}^{\textrm{UL}}$ is the FD vector of the transmitted symbols corresponding to bin $n$ defined as $\tilde{\mathbf{s}}_{:,1,n}^{\textrm{UL}} = [ \tilde{s}_{1,1,n}^{\textrm{UL}} \ \tilde{s}_{2,1,n}^{\textrm{UL}} \ \dots \ \tilde{s}_{K,1,n}^{\textrm{UL}} ]^{\textrm{T}}$.

Now that we have a compact expression for $\overline{\mathbf{y}}_{:,n}^{\textrm{UL}}$, we can apply the MRC-MMSE detector adapted from \cite{Kenney:TWC:2021}, which is given as
\begin{equation}
    \hat{\tilde{\mathbf{s}}}_{:,1,n}^{\textrm{UL}}= \alpha \sqrt{L} \left(  {\overline{\mathbf{A}}_n}^{\textrm{H}} \overline{\mathbf{A}}_n + L \sigma_w^2 \mathbf{I}_{K} \right)^{-1} {\overline{\mathbf{A}}_n}^{\textrm{H}} \overline{\mathbf{y}}_{:,n}^{\textrm{UL}}, 
    \label{eq:07}
\end{equation}
where $\alpha$ is the scaling factor that results in an unbiased estimate of the FD symbols.  Note that $\alpha$ is calculated for $ML$ antennas and $K$ UEs.  The factor of $\sqrt{L}$ before the inverse and the factor of $L$ within the inverse are due to the $1/\sqrt{L}$ scalar applied to $\overline{\mathbf{A}}_n$ in \eqref{eq:06}.  The estimates of the FD symbols are calculated for $N/L$ bins.  Afterward, the results are rearranged by UE to form $K$ length-$N/L$ vectors.  These FD vectors are then converted to the TD using a length-$N/L$ IFFT for each UE.

\subsection{Single-stream Performance Analysis} \label{Single_Stream_Analysis}
The performance of the MRC-MMSE detection for single-stream operation is characterized by the massive MIMO processing gain.  In \cite{Kenney:TWC:2021}, we defined the processing gain as the ratio of the output signal-to-interference-plus-noise ratio (SINR) to the input signal-to-noise ratio (SNR).  Given that the symbol rate of single-stream SCM is less than the noise bandwidth, we will define the processing gain as the ratio of output $E_s/N_0$ to input $E_s/N_0$ for each antenna, where $E_s/N_0$ is the energy per symbol divided by the noise power spectral density. At high input $E_s/N_0$, we see from \cite{Kenney:TWC:2021} that the value of $\alpha$ goes to unity.  Likewise, the $L \sigma_w^2 \mathbf{I}_{K}$ term in \eqref{eq:07} vanishes.  By examining the signal component of $\overline{\mathbf{y}}_{:,n}^{\textrm{UL}}$, we see that $(  \overline{\mathbf{A}}_n^{\textrm{H}} \overline{\mathbf{A}}_n )^{-1} \overline{\mathbf{A}}_n^{\textrm{H}} \overline{\mathbf{A}}_n \tilde{\mathbf{s}}_{:,n}^{\textrm{UL}} = \tilde{\mathbf{s}}_{:,n}^{\textrm{UL}}$ as desired.  The noise component of $\overline{\mathbf{y}}_{:,n}^{\textrm{UL}}$ is scaled by $\sqrt{L} ( \overline{\mathbf{A}}_n^{\textrm{H}} \overline{\mathbf{A}}_n )^{-1} \overline{\mathbf{A}}_n^{\textrm{H}}$.  The power of the noise scaling evaluated over all channel realizations is expressed as
\begin{equation}
    \begin{split}
	\frac{1}{K} \mathbb{E} \left[ \textrm{tr} \left\{ \left( \sqrt{L} \left( \overline{\mathbf{A}}_n^{\textrm{H}} \overline{\mathbf{A}}_n  \right)^{-1} \overline{\mathbf{A}}_n^{\textrm{H}} \right) \left( \sqrt{L} \left(  \overline{\mathbf{A}}_n^{\textrm{H}} \overline{\mathbf{A}}_n  \right)^{-1} \overline{\mathbf{A}}_n^{\textrm{H}} \right)^{\textrm{H}} \right\} \right] = \\
	\frac{L}{K} \mathbb{E} \left[ \textrm{tr} \left\{  \left( \overline{\mathbf{A}}_n^{\textrm{H}} \overline{\mathbf{A}}_n  \right)^{-1} \right\} \right], 
    \end{split}
    \label{eq:08}
\end{equation}
where the second line results from the fact that $( \overline{\mathbf{A}}_n^{\textrm{H}} \overline{\mathbf{A}}_n  )^{-1}$ is Hermitian symmetric.  From Lemma 6 of \cite{Verdu:2003}, we see that $\mathbb{E} [ \textrm{tr} \{ ( \overline{\mathbf{A}}_n^{\textrm{H}} \overline{\mathbf{A}}_n )^{-1} \} ]$ equals $K / (ML - K)$ since $ \overline{\mathbf{A}}_n^{\textrm{H}} \overline{\mathbf{A}}_n$ is a Wishart matrix.  The expression in \eqref{eq:08} reduces to $L/(ML-K)$, meaning that the resulting noise variance seen by each user will be $(M -( K/L ))$ times smaller than the input noise variance at each antenna, $\sigma^2_w$.  Hence, the processing gain for the UL single-stream (SS) detector is
\begin{equation}
	G^{\textrm{UL,SS}} = M -\frac{K}{L},
	\label{eq:08a}
\end{equation}
which is an improvement over the standard MRC-MMSE detector with processing gain of $M-K$, as shown in \cite{Kenney:TWC:2021}.

As a point of reference, we compare the single-stream SCM processing gain with that of DFT-precoded OFDM using an MMSE detector.  We assume that the same spectrum used for the SCM case is available for DFT-precoded OFDM (i.e., $N$ subcarriers).  For simplicity, we also assume that $N/L$ represents an integer number of resource blocks and $K/L$ is also an integer.  The $K$ UEs are divided into $L$ groups, where each group is assigned a mutually exclusive set of $N/L$ consecutive subcarriers on which to transmit.  At the BS, detection is performed for each group of $K/L$ UEs.  With $M$ BS antennas and $K/L$ users per group, the high-SNR processing gain is $M-(K/L)$.  This is identical to the single-stream SCM processing gain, showing that the average processing gain is equal for the two methods.  To continue the comparison, we acknowledge that the BS requires full-band CSI for each single-stream SCM UE.  In addition, the complexity of the calculations at the BS is greater for single-stream SCM than for DFT-precoded OFDM.  Despite these shortcomings, single-stream SCM may still be useful for some mMTC applications, given the arguments that follow.

It should be noted that the single-stream processing uses $ML$ virtual antennas, whereas in the DFT-precoded OFDM case, the detector uses $M$ antenna inputs.  As a result, the single-stream processing has better channel hardening, meaning that the variance of small-scale fading is reduced due to the channel being averaged over multiple antennas \cite{Bjornson:2016}.  Hence, single-stream processing has a smaller channel variance than DFT-precoded OFDM with the same number of physical antennas.  This implies that the number of physical antennas used for single-stream processing may be much smaller than what would normally be classified as massive MIMO, but the benefit of channel hardening will still be evident.

\subsection{Single-stream Simulation} \label{Single_Stream_Simulation}
\begin{figure}[!t]
    \centering
    \includegraphics[width=3.4in, clip=true, trim=3.0cm 10cm 3.75cm 10.5cm ]{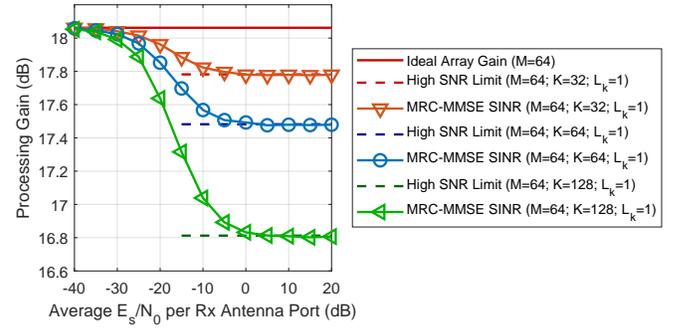}
    \caption{Single-stream simulation showing the normalized processing gain where $M=64$ and $L=8$.  Processing gain is shown for three values of $K$.  The high-$E_s/N_0$ limit is also plotted in each case based on \eqref{eq:08a}.}
    \label{single_stream_performance}
\end{figure}

To verify the high-$E_s/N_0$ asymptote for single-stream operation in \eqref{eq:08a}, we simulate a single-cell massive MIMO scenario with randomly selected channels.  It is assumed that each UE controls its received power at the BS such that the average received signal power per antenna is the same for all UEs (i.e., identical average input $E_s/N_0$).  The following parameters are used for the single-stream simulation: $M=64$ and $L=8$.  The number of UEs simulated are $32$, $64$, and $128$.  We calculate the processing gain versus the average $E_s/N_0$ at each antenna port.  Fig. \ref{single_stream_performance} shows that the processing gain closely approaches the $M-(K/L)$ level identified in the performance analysis.

\section{Multi-stream Detection} \label{Multi_Stream_Detection}
\begin{figure}[!t]
    \centering
    \includegraphics[width=3.4in, clip=true, trim=3.0cm 13.25cm 2.5cm 5cm ]{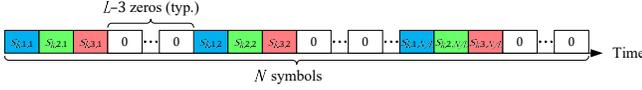}
    \caption{Multi-stream interleaving example for the $k^{\textrm{th}}$ UE with $L_k=3$.  Each stream consists of $N/L$ samples.}
    \label{multi_stream_interleaving}
\end{figure}

A more flexible alternative to the single-stream approach is to allow the $k^{\textrm{th}}$ user to transmit $L_k$ streams of data up to a maximum of $L$ streams.  Since each data stream is upsampled by $L$, the streams are interleaved together in the TD by circularly shifting one stream relative to another as depicted in Fig. \ref{multi_stream_interleaving}.  The value of the circular shift ranges from $0$ to $L-1$ and does not have any bearing on the performance.  For convenience, we assume that each additional stream of symbols from a given UE will be circularly shifted by one.  Accordingly, the transmitted TD vector from \eqref{eq:01} is modified for multiple streams as follows:
\begin{equation}
    \mathbf{x}_k^{\textrm{UL}} =  \mathbf{Z} \sum_{l=1}^{L_k} {\mathbf{E}_L}_{(l-1)} \mathbf{s}_{k,l}^{\textrm{UL}},
    \label{eq:09}
\end{equation}
where ${\mathbf{E}_L}_{(l-1)}$ is the expander matrix circularly shifted downward by $l-1$ and $\mathbf{s}_{k,l}^{\textrm{UL}}$ is the length-$N/L$ data vector from user $k$ corresponding to the $l^{\textrm{th}}$ stream.

When \eqref{eq:09} is converted to the FD, the cyclic shift of the expander matrix creates a new factor.  This can be visualized by realizing that the columns of the DFT matrix that are picked up by the expander matrix when converting to the FD will be different depending on the number of cyclic shifts.  Each successive column of the DFT matrix is obtained by multiplying the previous column by the second column of the DFT matrix.  We define $\omega=e^{-j 2 \pi / N}$ and create a vector $\bpsi = [ \omega^0 \  \omega^1 \ \dots \  \omega^{N-1} ]^{\textrm{T}}$ to represent the second column of the DFT matrix.  We define the diagonal matrix $\mathbf{\Psi}$ with diagonal elements that equal the elements of $\bpsi$ (i.e., $\textrm{diag}(\mathbf{\Psi}) = \bpsi$).  We can now express the product of the DFT matrix and the cyclically shifted expander matrix as
\begin{equation}
    \Fourier_N {\mathbf{E}_L}_{(l-1)} = \frac{1}{\sqrt{L}} \mathbf{\Psi}^{l-1} \mathbf{T}_{N,L} \Fourier_{N/L}.
    \label{eq:10}
\end{equation}

As in the case of the single-stream approach, each additional stream is spread across the entire bandwidth of the signal with the tiling matrix.  However, the copies of the spectrum have a different phase modulation due to the $\mathbf{\Psi}^{l-1}$ term in \eqref{eq:10}.  Consequently, it is necessary to represent the phase modulation when building up the composite matrix for each stream.  

\begin{figure*}[t]
\normalsize
\setcounter{temp_count}{\value{equation}}
\setcounter{equation}{11} 
\begin{equation}
	{\mathbf{B}}_{k,n} =
	\begin{bmatrix}
    	  \blambda_{:,k,n} & \omega^n \blambda_{:,k,n} & \omega^{2n} \blambda_{:,k,n} & \dots & \omega^{(L_k - 1) n} \blambda_{:,k,n} \  \\
    	 \blambda_{:,k,n+\frac{N}{L}} & \omega^{n+\frac{N}{L}} \blambda_{:,k,n+\frac{N}{L}} & \omega^{2(n+\frac{N}{L})} \blambda_{:,k,n+\frac{N}{L}} & \dots & \omega^{(L_k - 1) (n+\frac{N}{L})} \blambda_{:,k,n+\frac{N}{L}} \  \\
	  \vdots & \vdots & \vdots & \ddots & \vdots \\
    	  \blambda_{:,k,n+(L-1)\frac{N}{L}} & \omega^{n+(L-1)\frac{N}{L}} \blambda_{:,k,n+(L-1)\frac{N}{L}} & \omega^{2(n+(L-1)\frac{N}{L})} \blambda_{:,k,n+(L-1)\frac{N}{L}} & \dots & \omega^{(L_k - 1) (n+(L-1)\frac{N}{L})} \blambda_{:,k,n+(L-1)\frac{N}{L}} \  \\
	\end{bmatrix} 
	\label{eq:12}
\end{equation}
\setcounter{equation}{\value{temp_count} + 1}
\hrulefill
\vspace*{4pt}
\end{figure*}

Multi-stream detection supports flexible resource allocation by provisioning for each UE to send a number of data streams between $1$ and $L$ denoted by $L_k$ for UE $k$.  The values of $L_k$ need not be equal.  We begin specifying the detector by representing a column of $\mathbf{A}_n$ from \eqref{eq:05} as $\blambda_{:,k,n}$, which represents the channel coefficients for the $k^{\textrm{th}}$ UE at bin $n$ for each of the $M$ physical BS antennas.  We then define ${\mathbf{B}}_{k,n}$ as an intermediate matrix in \eqref{eq:12}, presented at the top of the next page, which is unique to the $k^{\textrm{th}}$ UE.  The matrix has dimensions $ML \times L_k$.  The $\blambda_{:,k,n}$ vectors corresponding to the same signal components (i.e., spaced by $N/L$ bins) are stacked vertically.  The first column does not have any additional phase modulation as in the single-stream case.  Each additional column has additional phase modulation, which changes for each stacked vector.  Although each column is based on the same set of $\blambda_{:,k,n}$ vectors, the variation in phase keeps the matrix's rank full. The ${\mathbf{B}}_{k,n}$ matrices from all the UEs are combined to form the multi-stream version of the composite channel matrix $\overline{\mathbf{B}}_n = [ {\mathbf{B}}_{1,n} \ {\mathbf{B}}_{2,n} \ \dots {\mathbf{B}}_{K,n} ] / \sqrt{L}$.  Note that a factor of $1/\sqrt{L}$ was included in $\overline{\mathbf{B}}_n$ for the MSP case.

Next, we define the structure of the composite vector of FD symbols that corresponds with $\overline{\mathbf{B}}_n$.  For each UE, we define the vector of FD symbols for the $n^{\textrm{th}}$ bin as $\tilde{\mathbf{s}}_{k,:,n}^{\textrm{UL}} = [ \tilde{s}_{k,1,n}^{\textrm{UL}} \ \tilde{s}_{k,2,n}^{\textrm{UL}} \ \dots \tilde{s}_{k,L_k,n}^{\textrm{UL}} ]^{\textrm{T}}$.  These intermediate vectors are of length $L_k$, which could vary from UE to UE.  The composite FD symbol vector is now defined as $\overline{\mathbf{s}}_n^{\textrm{UL}} = [ ( \tilde{\mathbf{s}}_{1,:,n}^{\textrm{UL}} )^{\textrm{T}} \ ( \tilde{\mathbf{s}}_{2,:,n}^{\textrm{UL}} )^{\textrm{T}} \dots ( \tilde{\mathbf{s}}_{K,:,n}^{\textrm{UL}} )^{\textrm{T}} ]^{\textrm{T}}$.  The composite vector length is $K_{\textrm{v}} = \sum_{k=1}^{K} L_k$, which is the number of virtual users.


Taking the $\overline{\mathbf{B}}_n$ matrix in place of $\overline{\mathbf{A}}_n$ and $\overline{\mathbf{s}}_n^{\textrm{UL}}$ in place of $\tilde{\mathbf{s}}_{:,n}^{\textrm{UL}}$, we use the same structure as \eqref{eq:06} to form the multi-stream expression for the FD composite received vector.  We can apply the MRC-MMSE detector similar to \eqref{eq:07} yielding
\begin{equation}
    \hat{\overline{\mathbf{s}}}_{n}^{\textrm{UL}}= \alpha \left(  \overline{\mathbf{B}}_n^{\textrm{H}} \overline{\mathbf{B}}_n + \sigma_w^2 \mathbf{I}_{K_{\textrm{v}}} \right)^{-1} \overline{\mathbf{B}}_n^{\textrm{H}} \overline{\mathbf{y}}_{:,n}^{\textrm{UL}}, 
    \label{eq:12}
\end{equation}
where $\alpha$ is the scaling factor calculated for $ML$ antennas and $K_{\textrm{v}}$ UEs.  The estimates of the FD symbols are calculated for the first $N/L$ bins.  After the estimates for the $N/L$ bins have been calculated, the results are rearranged to form $K_{\textrm{v}}$ length-$N/L$ vectors corresponding to the FD estimates of the symbol vectors for each stream.  The FD vectors are then converted to the TD using the IFFT.

\subsection{Multi-stream Performance Analysis} \label{Multi_Performance_Analysis}

The multi-stream performance analysis is very similar to the single-stream analysis.  The major difference is that the number of virtual users, $ K_{\textrm{v}}$, replaces the $K$ users in the single-stream case.  It is also important to note that the presence of multiple streams for a given user means that the rows of $\mathbf{B}_{k,n}$, and by extension $\overline{\mathbf{B}}_n$, are not Gaussian vectors.  Even though $\overline{\mathbf{B}}_n^{\textrm{H}} \overline{\mathbf{B}}_n$ is not a Wishart matrix per Definition 2.3 of \cite{Verdu:Random_Matrix_Theory}, the columns of $\overline{\mathbf{B}}_n$ are still uncorrelated due to the variation of the phase shifts.  We find that $\mathbb{E} [ \textrm{tr} \{ ( \overline{\mathbf{B}}_n^{\textrm{H}} \overline{\mathbf{B}}_n )^{-1} \} ]$ equals $K/(ML-K_{\textrm{v}})$ as though $\overline{\mathbf{B}}_n^{\textrm{H}} \overline{\mathbf{B}}_n$ was a Wishart matrix.  It follows that each additional stream functions the same as an additional single-stream user.  Consequently, the multi-stream (MS) processing gain for the UL is
\begin{equation}
	G^{\textrm{UL,MS}} = M -  \frac{ K_{\textrm{v}} }{L}.  \label{eq:last}
\end{equation}
For example, if $M=64$, $K=32$, $L = 4$, and $L_k = 3$ for all $k$, then the processing gain would be approximately $16$ dB.  Since $K<M$ in the example, we could use the MRC-MMSE detector from \cite{Kenney:TWC:2021} with processing gain equal to $M-K$, but it would yield a gain of only $15$ dB, resulting in a $1$ dB penalty.

As previously mentioned, the matrix $\overline{\mathbf{B}}_n^{\textrm{H}} \overline{\mathbf{B}}_n$ is full rank and therefore invertible.  We note that even though portions of the channel coefficients are duplicated in the composite channel matrix as shown in \eqref{eq:12}, the condition number of $\overline{\mathbf{B}}_n^{\textrm{H}} \overline{\mathbf{B}}_n$ is not degraded.  Simulation results show that the condition number of $\overline{\mathbf{B}}_n^{\textrm{H}} \overline{\mathbf{B}}_n$ actually improves (i.e., lower value) when compared to that of $\mathbf{A}_n^{\textrm{H}} \mathbf{A}_n$ for all cases where $L_k < L$.

\subsection{Multi-stream Simulation Results} \label{Multi_Simulation_Results}

\begin{figure}[!t]
    \centering
    \includegraphics[width=3.4in, clip=true, trim=3.0cm 10cm 3.75cm 10.5cm ]{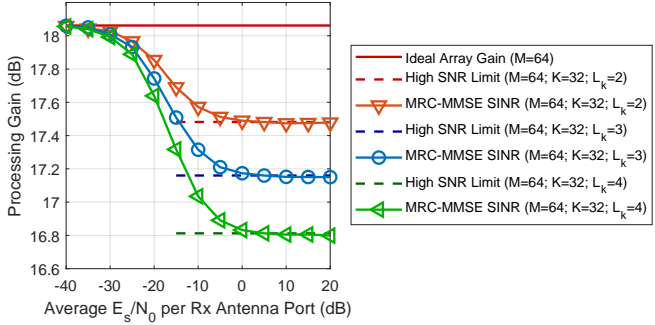}
    \caption{Multi-stream simulation where $M=64$, $L=8$, and $K = 32$.  The number of streams is $L_k=2$, $L_k=3$, and $L_k=4$ for all $k$.  The high-$E_s/N_0$ limit is also plotted in each case based on \eqref{eq:last}.  Note that the multi-stream processing gain with $K_{\textrm{v}}$ streams is identical to that of single-stream operation when the number of single-stream users equals $K_{\textrm{v}}$.}
    \label{multi_stream_performance}
\end{figure}

The multi-stream simulation follows the same scenario as the single-stream scenario, but $K$ is fixed at $32$ users.  This is the smallest value used in the single-stream simulation.  The number of streams is constant for all users.  Here we set $L_k=2$, $L_k=3$, and $L_k=4$ for all users.  The results are shown in Fig. \ref{multi_stream_performance}.  Notice that the number of virtual users, $L_k K$, are $64$, $96$, and $128$.  The first and the last of these values match the number of virtual users for two of the single-stream cases shown in Fig. \ref{single_stream_performance}.  As predicted in \eqref{eq:last}, the high-$E_s/N_0$ gain matches, showing that the processing gain is dependent upon the total number of virtual users.

\section{Conclusion} \label{Conclusion}
We have presented a means of detecting SCM waveforms in the FD in a flexible manner that allows for a large number of simultaneous UEs.  Specifically, the number of UEs can exceed the number of BS antennas, and each UE can transmit using the number of data streams dictated by the BS.  The performance achieved by the single- and multi-stream configurations was analyzed to produce asymptotic bounds at high-$E_s/N_0$ for the massive MIMO processing gain.  The processing architecture allows for resource allocation flexibility similar to resource blocks used in 4G LTE and 5G NR.  The MSP algorithm also benefits from the massive MIMO effect of channel hardening with a lower physical antenna count.  Another advantage to the SCM approach is that the signal is spread over all of the frequency bins, so each symbol sees the average channel condition.


\end{document}